\documentclass[pra,preprint,aps,showpacs]{revtex4}

\usepackage{amssymb,amsmath}
\usepackage{graphicx}

\begin{document}
\title{On Exchange of Orbital Angular Momentum Between Twisted Photons and Atomic Electrons}

\date{\today} 

\author{Basil S. Davis}
\email{bdavis2@tulane.edu}
\author{L. Kaplan}
\email{lkaplan@tulane.edu}
\author{J. H. McGuire}
\email{mcguire@tulane.edu}
\affiliation{Department of Physics and Engineering Physics,
Tulane University, New Orleans, LA 70118}


\begin{abstract}
We obtain an expression for the matrix element for a twisted (Laguerre-Gaussian profile) photon scattering from a hydrogen atom. We consider photons incoming with an orbital angular momentum (OAM) of $\ell \hbar$, carried by a factor of $e^{i \ell \phi}$ not present in a plane-wave or pure Gaussian profile beam. The nature of the transfer of $+2\ell$ units of OAM from the photon to the azimuthal atomic quantum number of the atom is investigated. We obtain simple formulae for these OAM flip transitions for elastic forward scattering of twisted photons when the photon wavelength $\lambda$ is large compared with the atomic target size $a$, and small compared the Rayleigh range $z_R$, which characterizes the collimation length of the twisted photon beam.
\end{abstract}

\maketitle

\section{Introduction}
\label{sec:intro}

The interaction of twisted photons with atomic electrons differs from that of plane-wave photons.  This difference arises from a phase factor of $e^{i \ell \phi}$, which changes the parity symmetry enabling the twisted photon to transfer orbital angular momentum (OAM) to a hydrogen atom. In addition, the vortex geometry of a twisted photon may allow more flexibility in how it interacts with an atom than does the simpler symmetry of a plane-wave photon.  While a plane wave photon is characterized only by its wavelength, a twisted photon, like a Gaussian beam, possesses an additional length scale, namely the waist size of the vortex.  An atom typically offers possible variation of quantum numbers $N$, $L$, and $M$ (ignoring the usually weak spin flip of the electron itself) to entice a photon to react.  The atomic size varies with $N$, while $L$ and $M$ regulate the transfer of angular momentum.  Twisted photons can transfer OAM to an atom in addition to the spin angular momentum (SAM) that may be exchanged by a plane-wave (Gaussian profile) photon.

The existence of OAM in a twisted photon, as distinct from its SAM, has been established on classical and quantum mechanical grounds \cite{PoyntingOAM,Beth,Barnett, Humblet,
Molina-TerrizaTorresTorner,BabikerBADR}.  A collimated coherent beam of photons that carry OAM is called an optical vortex \cite{MolinaTerrizaRTTW}, a twisted beam \cite{BerryMcDonald}, or a helical laser beam \cite{GiovanniniNMS}. Such beams have a Laguerre-Gaussian transverse profile. Whereas there is an analogy between photonic and atomic orbital angular momenta, there is a significant difference. An atomic state of total angular momentum $L$ is $(2L+1)$-fold degenerate, as there are $2L +1$ possible values of the corresponding magnetic quantum number $M$. By contrast, a photonic state of orbital angular momentum $|\ell|$ has a two-fold degeneracy, with the ``magnetic" quantum number $m$ taking on one of the two values $+\ell$ and $-\ell$. Any such pair of modes with opposite OAM values and a common radial profile defines a basis for a two-dimensional OAM subspace in which one can encode a generic qubit \cite{GiovanniniNMS}.

The two basic single-qubit flip operations in quantum computation are the bit flip and the phase flip. Both of these operations have been implemented successfully with photons possessing orbital angular momentum \cite{PadgettCourtial,MolinaTerrizaRTTW}. Because there is the possibility of an unwanted interaction between the photons and the atoms of the medium involving an exchange of angular momentum, errors can interfere with the communication of quantum information using such twisted photons. The interactions of photons with atoms have been studied for nearly a century. But the interaction of twisted photons with matter has been the focus of attention only within the past decade or so \cite{SimulaNHCSM,BabikerBADR,AllenBLL},
even though OAM transfer from a twisted photon to an atomic electron has not been described during that time.
Elastic scattering of twisted photons into the forward direction with no change in their
radial distribution is of special interest in quantum computing \cite{BarreiroWeiKwiat,WildeUskov,Uskov}, motivating this paper.

Here we obtain, for the first time, an expression for the dominant matrix element for a transition from $\ell$ to $-\ell$ for elastic forward scattering of twisted photons interacting with atomic hydrogen. We develop an explicit expression for, and analyze, this matrix element. Whereas a flip of the SAM of a plane-wave photon scattered elastically forward by an atom is forbidden \cite{Heitler}, we show that such a restriction does not hold for an OAM flip for a twisted photon. For twisted photons elastically scattered into the forward direction, we show that a significant transition rate may be possible for this (OAM) process in atomic hydrogen, especially when the photon wavelength $\lambda$ is not much larger than the target size $a$ and not much smaller than $z_R$, the Rayleigh range that sets the geometry of the optical vortex.

In what follows we first identify the matrix element of interest in this process, and specify our assumptions and approximations.
Then we formulate elastic forward scattering of photons from atomic hydrogen for plane-wave photons, noting that a transition from $s = \pm 1$ to $s = \mp 1$ is forbidden for such plane-wave photons. Next we replace the incoming and outgoing plane-wave photons by twisted photons, which carry OAM $\ell$ in addition to the SAM $s$. The transition rate for twisted photons elastically scattered into the forward direction is non-zero and may be significant. We subsequently discuss the physical nature of this $\ell \rightarrow -\ell$ transition,  interactions with macroscopic targets, and possible future directions and applications. We conclude with a summary of the main points of this paper.

In this paper we generally work in atomic units $e^2 = m_e = \hbar = 1$, except where explicit inclusion of $\hbar$ or $m$ is helpful in clarifying a concept.

\section{Central matrix element; assumptions and approximations}
\label{matrix}

\subsection{Our central matrix element}

In the wave picture, the first Born matrix element for electronic transitions in atoms
interacting with plane-wave photons is
\begin{equation}
  M = \langle f|  e^{-i \vec{k}_f \cdot \vec{r}}  e^{ i \vec{k}_i \cdot \vec{r}}  |i\rangle 
= \langle f|  e^{i \vec{q} \cdot \vec{r}} |i\rangle   
=  \int \    \phi_f^*(\vec{r}) e^{i \vec{q} \cdot \vec{r}} \phi_i(\vec{r})  \  d\vec{r} 
= \langle F|I\rangle / N_\gamma^2 \,,
\label{I}
\end{equation}
where $(E_f-E_i)|M|^2/|\vec{q}|^2$ is the generalized oscillator strength~\cite{Heitler, MC1,McGuiregos}.
This matrix element, $ M = \langle f|  e^{i \vec{q} \cdot \vec{r}} |i\rangle$, 
is the key quantity of interest in this paper. 
Here, $|i\rangle$ and $|f\rangle$  correspond to initial- and final-state atomic wave functions $\phi_i$ and $\phi_f$ with energies $E_i$ and $E_f$, and  $\vec{q} = \vec{k}_i - \vec{k}_f$  is the momentum transferred by the photon, where $\vec{k}_i$ and $\vec{k}_f$ are the momenta of the initial (incoming) and final (outgoing) photons. We note that the matrix element $M$ for photon scattering (considered here) is the same as that for photo-absorption or emission (not evaluated in this paper) of a photon of momentum $\vec{q}$. 
In Eq.~(\ref{I}),  $\langle F|I\rangle$ is the projection of the properly normalized initial state of the system onto the final state.
The factor $N_\gamma$ appears in the volume normalization of the electromagnetic field,
\cite{McGuire0}, $\vec{A} = \sqrt{2 \pi c^2/\omega V} \hat{\Lambda} e^{i(\vec{k} \cdot \vec{r} - \omega t) } = \hat{\Lambda} N_\gamma e^{i (\vec{k} \cdot \vec{r} -\omega t)} $,
where $\hat{\Lambda}$ is the unit vector in the direction of the polarization, $\omega$ is the photon frequency, and $V$ is the spatial volume of the photon. 
In the special case of  elastic scattering with scattering angle $\Theta$,  $q = k \sin\Theta$, where $k_i = k_f = k$.
If $\vec{q}$ is chosen to be along the $z$ axis, expanding $e^{i \vec{q} \cdot \vec{r}} = e^{i q z}$ yields a series expansion for $M$ 
in dipole matrix elements~\cite{MC2}, namely,
\begin{equation}
(q z_{if})^n = q^n \int \;   \phi_f^*(\vec{r}) ( r \cos\theta)^n  \phi_i(\vec{r})  \;  d\vec{r} \,.
\label{znif}
\end{equation}
We note that for visible photons interacting with electrons in low-lying atomic energy levels, 
$qa \leq 2 \pi a/\lambda \ll 1$, where $a$ is the radius of our hydrogen atom. 
Via the $e^{i \vec{q} \cdot \vec{r}}$ factor,  
this reduces the cross section from $\pi a^2$ by a factor of order 
$(a/\lambda)^2 \sim 10^{-8} $, i.e., from the geometric cross section of the atom to the smaller Thompson scattering 
cross section, as noted below.

Photo-absorption (or emission) differs from scattering of a photon.  In the former case there is a single incoming (or outgoing) photon, and the direction of the photon momentum, $\vec{k}$, is a natural axis of symmetry. Transfer of angular momentum must take place along the $\hat{k}$ direction in photo-absorption, corresponding to the selection rule $\Delta s = \pm 1$ for SAM transfer by plane-wave photons.

By contrast, for photon scattering, of interest to us in this paper, there are both incoming and outgoing photons, 
in general with different momenta, $\vec{k}_i$ and $\vec{k}_f$.
Now, as we see in Eq.~(\ref{I}), it is the momentum transfer $\vec{q}$ that determines the axis of symmetry. Furthermore, it is $\vec{q}$ that corresponds to the momentum $\vec{P}$ in the classical equation for the electronic orbital angular momentum, $\vec{\cal{L}} = \vec{R} \times \vec{P}$.
This yields the selection rule $\Delta s = \pm 1$, corresponding to transfer of one unit of $s$ {\it parallel to $\vec{q}$}. In the case of inelastic scattering, $q \geq q_{\rm min} > 0$ 
and $\hat{q} \neq \hat{k}$ except for scattering angle $\Theta = 0$, i.e., except for perfectly forward scattering.

For elastic scattering, $k_i = k_f = k$ and $\vec{q} =  \hat{q} \ k \sin\Theta$,
where $\hat{q}\cdot \hat{k}= \sin\Theta \rightarrow 0$ as $\Theta \rightarrow 0$.  
That is, for elastic scattering of photons \cite{Glauber56} 
(or incident plane-wave protons or electrons \cite{MC1}), 
$\hat{q}$ serves as a mathematical axis of symmetry, and near the limit $\Theta \rightarrow 0$,
there are two mutually perpendicular axes of symmetry, namely $\hat{k}$ and $\hat{q}$.

\subsection{Assumptions and approximations}
\label{sec:assumptions}

In evaluating the matrix element for scattering of twisted photons from atomic hydrogen in section IV below, we use a number of assumptions and approximations.

The coupling $H_{\rm int}$ between an atomic electron and a photon arises from the kinetic energy term in the Hamiltonian of the photon-atom system when 
$ P^2/2m \rightarrow (\vec{P} - \frac{e}{c}\vec{A})^2/2m = P^2/2m + H_{\rm int}$. 
The scattering amplitude, containing our matrix element $M$, is conventionally expanded in a perturbation series in $H_{\rm int}$,
with the lowest-order term consisting of $\vec{P} \cdot \vec{A}$, $\vec{A} \cdot \vec{P}$, 
and $\vec{A} \cdot \vec{A}$ terms.  
The first two terms correspond to photo-absorption or photo-emission~\cite{MairVWZ}, and the last term represents the scattering of the photon by the electron~\cite{Heitler}.  
For plane-wave photons in the visible part of the electromagnetic spectrum, photo-absorption is normally the dominant process. But it is not of interest to us here, except to explain, e.g. below Eq.~(\ref{I}), how photon absorption differs from photon scattering.
Thus we focus on the $\vec{A} \cdot \vec{A}$ term. 

We have also considered the contribution of second-order terms in $\vec{P} \cdot \vec{A}$ 
and $\vec{A} \cdot \vec{P}$ to photon-electron scattering. Working in the length gauge, where we may avoid the non-commutativity of the momentum
$\vec{P} = -i\hbar \nabla_{\vec{r}}$ with other operators, one finds that 
these second-order terms give a contribution of order $(a/\lambda)^2$ compared
to the first-order $\vec{A} \cdot \vec{A}$ term, since each $P$ operator gives rise to a factor of $a/\lambda$, assumed to be small.  For these $(\vec{P} \cdot \vec{A})^2$ type terms, propagation in intermediate states is suppressed by additional factors $a/\lambda$.
Also we neglect the small momentum coupling with the atomic nucleus~\cite{{BabikerBADR}}, i.e., the translational degree of freedom of the atom is not included.   

We consider scattering from atomic hydrogen.  Not all matter is hydrogenic.  One limitation is that the hydrogenic angular momentum quantum numbers are not exact for other atoms \cite{Lin}.
Also atoms could be externally aligned, or could have different polarizations.  As usual, we assume that the spin flip of an atomic electron, due to a magnetic field, is rare for photon scattering. 

For scattering by a twisted photon, we assume that the atom is at the center of the optical vortex.  Obviously, we make this assumption for the sake of mathematical symmetry rather than experimental convenience.  We discuss this later. 

Thus, in essence we assume that we may simply redo the matrix calculation for plane-wave photons (Eq.~(\ref{I})), replacing the plane wave fields by those for twisted photons.

\section{Matrix element for plane-wave photons}
\label{pwphotons}

The matrix element arising from the $\vec{A}\cdot \vec{A}$ term, where $\vec{A}$ corresponds to a plane-wave photon, has been evaluated in atomic physics in processes related to Compton scattering.  The relevant matrix element is~\cite{McGuire1,Heitler}  
\begin{equation}
M_{\rm Compton} = (\hat{\Lambda}_f \cdot \hat{\Lambda}_i)M= (\hat{\Lambda}_f \cdot \hat{\Lambda}_i) \langle f|e^{i \vec{q} \cdot \vec{r}}|i\rangle 
= (\hat{\Lambda}_f \cdot \hat{\Lambda}_i) \langle f|e^{-i \vec{k}_f \cdot \vec{r}} e^{i \vec{k}_i \cdot \vec{r}}|i\rangle \,.
\label{IA2}
\end{equation}
Thus $\hat{\Lambda}_f \cdot \hat{\Lambda}_i \rightarrow 1$, i.e., change of polarization is forbidden. 
Furthermore, for elastically scattered photons in the forward direction ($q \rightarrow 0$), this matrix element goes to zero when $|f\rangle \neq |i\rangle$
since $\langle f|i\rangle = 0$ for the atomic wave functions.  

\section{Matrix element for twisted photons}

\subsection{Spatial dependence of a twisted photon}

The vector potential for a twisted photon propagating in the $z$ direction \cite{AllenBSW} may be written as
\begin{equation} 
\vec{A} ={\hat{\Lambda}}\,u(\rho,z,\phi) e^{ik z} \,,
\end{equation}
where the polarization direction $\hat{\Lambda}$ lies in the $x$-$y$ plane.
For a specific mode characterized by radial index $p$ and winding number $\ell$, the transverse profile in the paraxial approximation is given by~\cite{AllenBSW, YaoPadgett},
\begin{eqnarray} u_{p, \ell}(\rho,z,\phi) &=& 
  \sqrt{\frac{2p!}{\pi (p+|\ell|)!}} \, \frac{1}{w(z)} \left [ \frac{\rho\sqrt{2}}{w(z)}\right ]^{|\ell|}\exp\left [ -\frac{\rho^2}{w^2(z)} \right ]L_{p}^{|\ell|}\left ( \frac{2\rho^2}{w^2(z)} \right ) \nonumber \\
 &\times& \exp[i\ell\phi]\,\exp\left [ \frac{ik \rho^2z}{2(z^2+z_R^2)} \right ] \exp\left [ -i(2p +|\ell| +1)\tan^{-1} \left ( \frac{z}{z_R} \right )\right ]  \,. \label{LG1}\end{eqnarray}
Here the photon carries momentum $\vec{k}$ propagating in the $\hat{z}$ direction,
$w(z) = w(0)\sqrt{1+z^2/z^2_R}$, and
$w(0) = \sqrt{\lambda z_R/\pi}$ is a measure of the minimum width of the beam (beam waist) determined by the Rayleigh range, $z_R$ .  
The phase $(2p + |\ell| +1) \tan^{-1}({z}/{z_R})$ is the Gouy phase, and $ L_p^{|\ell|}(x)$ is an associated Laguerre  polynomial related to the more familiar Laguerre polynomials by 
$ L_p^{|\ell|}(x) = (-1)^{|\ell|} \frac{{d}^{|\ell|}}{dx^{|\ell|}} L_{p+|\ell|}(x)  $. 
The index $p$ is the number of radial nodes in the intensity distribution, and $\ell$ is the azimuthal index, which is of special interest in this paper.  

We now choose to re-express the spatial function of the twisted photon in spherical coordinates, taking $\rho= r \sin \theta$ and $z = r \cos \theta$, in order to calculate matrix elements.   Then we have,
\begin{eqnarray} & & u_{p, \ell}(r,\theta,\phi) = 
 \sqrt{\frac{2p!}{\pi (p + |\ell|)!}} \frac{1}{w(r \cos\theta)}\left [\frac{r|\sin\theta|\sqrt{2}}{w(r \cos\theta)}\right]^{|\ell|} \exp\left[- \frac{(r\sin\theta)^2}{w^{2}(r \cos\theta)}\right] \exp(i \ell \phi) \nonumber \\
&\times &L_p^{|\ell|}\left(\frac{2 (r\sin\theta)^2}{w^{2}(r \cos\theta)} \right)
\exp\left [ \frac{ ik (r\sin\theta)^2 r\cos\theta}{2((r \cos\theta) ^2 + z^2_R)}\right] \exp\left[-i(2p + |\ell| +1) \tan^{-1}\left(\frac{ r \cos\theta } {z_R}\right)\right] \,.\label{sphercoord} \end{eqnarray}
The factor that is the key to transfer of OAM is $e^{i \ell \phi}$,
which can couple with a mathematically similar factor in the wave function of a hydrogen atom.
This $e^{i \ell \phi}$ phase factor is not present in the spatial distribution for a plane-wave photon or Gaussian beam.
We note that the Gouy phase depends on $|\ell|$ and not $\ell$.  
This means that the parity of the Gouy phase does not change as $\ell \rightarrow -\ell$.
This avoids an additional mirror antisymmetry that can occur when $\psi \rightarrow \psi^*$.  
This will be key to understanding the parity of the photon-atom system, which determines
how the quantum number $\ell$ of the twisted photon is allowed to change to $-\ell$, with a corresponding change $\Delta M = 2\ell$ in the orbital angular momentum of the hydrogen atom. 

\subsection{Derivation of the matrix elements for twisted photons}

Now we apply the same assumptions used to obtain Eq.~({\ref{IA2}) for the $\vec{A}\cdot \vec{A}$ matrix element for plane-wave photons, replacing the expression $e^{i \vec{k} \cdot \vec{r}}$ 
in $\vec{A}$ in Eq.~(\ref{IA2})  by the corresponding, more complex, expression from 
Eq.~(\ref{sphercoord}) for twisted photons.  

For scattering away from the forward direction, the initial and final photon states 
acquire different momentum directions $\vec{k}$.  Often it may be advantageous to choose 
the $z$-axis parallel to $\vec{q} = \vec{k}_i - \vec{k}_f$, as noted below Eq.~(\ref{znif}).
However, we will soon be interested in elastic scattering in the forward direction where 
$\vec{q} = 0$.  
For this reason it is convenient here to choose the $z$-axis parallel to $\hat{k}_i$
to preserve the mathematical form of Eq.~(\ref{sphercoord}), and to let $M_{i,f}$ be the initial
and final $z$-components of the electronic orbital angular momentum.
Thus, taking $u_{p,\ell}$ from Eq.~(\ref{sphercoord}), we have,
\begin{equation}
M  =  \int d\vec{r} \,  \phi_{N_f,L_f,M_f}^*(\vec{r}) e^{-i\vec{k}_f \cdot \vec{r} } 
e^{i\vec{k}_i \cdot \vec{r} } u_{p_f,  \ell_f}^*(\vec{r}) \  
 u_{p_i, \ell_i}(\vec{r}) \phi_{N_i,L_i,M_i}(\vec{r})    \nonumber
\end{equation}
\begin{equation}
=   \int d\vec{r}  \,   \phi_{N_f,L_f,M_f}^* (\vec{r})       
 \sqrt{\frac{2p_f!}{\pi (p_f + |\ell_f |)!}} \frac{1}{w(r \cos\theta')} 
 \left[\frac{ r|\sin \theta'| \sqrt{2}}{w(r \cos\theta')}\right]^{|\ell_f | } 
\exp\left[ \frac{-(r \sin\theta')^2}{w^{2}(r \cos\theta')}\right]\nonumber\end{equation}
\begin{equation}\times L_{p_f}^{|\ell_f |}\left(\frac{2 (r \sin\theta')^2}{w^{2}(r \cos\theta')}\right) 
\exp(-i l_f \phi')
\exp\left[ \frac{-i k_f (r \sin\theta' )^2 r\cos\theta'}{2((r \cos\theta') ^2 + z^{2}_R)}\right] 
\exp\left[i(2p_f + |\ell_f | +1) \tan^{-1}\left(\frac{ r \cos\theta' }{z_R}\right)\right]    
 \nonumber
 \end{equation} 
 \begin{equation} \times  e^{i\vec{q} \cdot \vec{r} }
 \sqrt{\frac{2p_i!}{\pi (p_i + |\ell_i |)!}} \frac{1}{w(r \cos\theta)} 
    \left[\frac{ r|\sin \theta| \sqrt{2}}{w(r \cos\theta)}\right]^{|\ell_i |} 
\exp\left[ \frac{-(r \sin\theta)^2}{w^{2}(r \cos\theta)}\right]
L_{p_i}^{|\ell_i |}\left(\frac{2 (r \sin\theta)^2}{w^{2}(r \cos\theta)}\right) \nonumber\end{equation}
\begin{equation}\times  \exp(i \ell_i \phi)
\exp\left[ \frac{ ik_i (r \sin\theta )^2 r\cos\theta}{2((r \cos\theta) ^2 + z^{2}_R)}\right] 
\exp\left[-i(2p_i + |\ell_i | +1) \tan^{-1}\left(\frac{ r \cos\theta }{z_R}\right)\right]  
    \phi_{N_i,L_i,M_i} (\vec{r}) 
    \label{Mt2}
\end{equation}
Here ($\theta$, $\phi$) are spherical angles defined so that the north pole $\theta=0$ corresponds to the $\hat{k}_i$ direction, while
$(\theta'$, $\phi')$ is a coordinate system rotated by the scattering angle $\Theta$, so that $\theta'=0$ corresponds to the $\hat{k}_f$ direction. As in Eq.~(\ref{IA2}), $M$ is multiplied in the full matrix element by the 
polarization factor $\hat{\Lambda}_f \cdot \hat{\Lambda}_i$.
Eq.~(\ref{Mt2}) is the most general expression in this paper.  It is valid for all scattering angles, 
not just $\Theta =0$ that is one focus of attention in this paper.

For applications in quantum computation cited in the introduction, and for mathematical simplicity,
we focus attention on cases where
$\ell_f = -\ell_i = -\ell$, corresponding to an OAM qubit flip of the twisted photon.    
In the case of forward scattering, this implies a transfer of precisely $2\ell\hbar$ units of angular momentum.
We also take $p_f = p_i = p$ and $N_f=N_i=N$, causing the Gouy phase to disappear from the matrix element.
Then,
\begin{equation}
M =  \frac{ 2p!}{\pi (p + |\ell|)!}   
\int_0^\infty r^2 dr \int_0^\pi \sin(\theta) d\theta  \int_0^{2\pi} d\phi \   \  \phi_{N,L_f,M_f}^* (r,\theta,\phi)       
 \frac{1}{(w(r \cos\theta))^2} 
\nonumber\end{equation}
\begin{equation}\times \left[\frac{ r|\sin \theta| \sqrt{2}}{w(r \cos\theta)}\right]^{2|\ell| } 
\exp\left[ \frac{-2(r \sin\theta)^2}{w^{2}(r \cos\theta)}\right] 
\left( L_p^{|\ell |}\left(\frac{2 (r \sin\theta)^2}{w^{2}(r \cos\theta)}\right) \right)^2
e^{i \vec{q} \cdot \vec{r} } 
\exp\left[ \frac{i \vec{q} \cdot \vec{r} (r \sin\theta )^2}{2((r \cos\theta) ^2 + z^{2}_R)}\right] 
\nonumber\end{equation}
\begin{equation}\times  
e^{i 2 \ell \phi}   \phi_{N,L_i,M_i} (r,\theta,\phi)  \,.
\label{Mt2p}
\end{equation}
In the integration over $\phi$, an allowed transition occurs when 
$M_f - M_i  = 2 \ell$, so that $2 \ell$ units of OAM
are transferred from the twisted photon to the hydrogenic target. 
This requires that the principal quantum number of the target atom 
be restricted to $N \geq N_{\rm min}$, where $N_{\rm min} = |\ell| + 1$. 

For scattering away from the forward direction,
when $\lambda \gg a$, rapidly oscillating $e^{i\vec{q} \cdot \vec{r}}$ terms 
may reduce transition rates.  
When $\vec{k}_f \neq \vec{k}_i$, it is advantageous to choose 
the $z$-axis parallel to $\vec{q} = \vec{k}_i - \vec{k}_f$, as noted below Eq.~(\ref{znif}),
since the calculation simplifies and selection rules emerge. 
We also note that the matrix element of Eq.~(\ref{Mt2p}) reduces to the matrix element for Gaussian beams, 
when $\ell \rightarrow 0$ and $p \rightarrow 0$, and this expression in turn reduces to the matrix element for plane waves in the limit
$w(0) \rightarrow \infty$ and hence $z_R \rightarrow \infty$. 

\subsection{Matrix element to leading order in $a/w(0)$}  

In this section we restrict our attention to physical conditions where $N_i = N_f = N$ and $a \ll w(0)$.
For  example, $a/w(0) \simeq 10^{-7}$ when the beam width of the order of a millimeter and $a=a_0$ is the radius of a hydrogen atom.
Under such conditions it makes sense to expand the matrix element in Eq.~(\ref{Mt2p}) 
in powers of $a/w(0)$.
Noting that $L_p^{|\ell|}(x) \rightarrow \binom {p + |\ell|}{p}  =     \frac{(p + |\ell|)!}{p!  |\ell|!}  $ as $x \rightarrow 0$, we find that, to lowest order, the matrix element $M$ reduces to 
\begin{equation}
M \simeq   \frac{2p!}{\pi (p + |\ell  |)!}       \left(  \frac{(p + |\ell|)!}{p!  |\ell|!}  \right)^2
 \int_0^\infty r^2 dr \int_0^\pi \sin(\theta) d\theta  \int_0^{2\pi} d\phi \   \  \phi_{N,L_f,M_f}^* (r,\theta,\phi)        
\nonumber\end{equation}
\begin{equation}\times 
 \frac{1}{(w(r \cos\theta))^2}  
\left[\frac{ r|\sin \theta| \sqrt{2}}{w(r \cos\theta)}\right]^{2|\ell | } 
e^{i 2 \ell \phi}    \phi_{N,L_i,M_i} (r,\theta,\phi)  \,.
\label{Mt2a}
\end{equation}
This is obviously non-zero, so that an OAM angular momentum exchange of $2\ell \hbar$ 
can occur for twisted photons elastically scattered into the forward direction.
It is also apparent that the transition rate scales with $|\ell|$ as $(a/w(0))^{4|\ell|}$, due to the reduction of the twisted beam intensity along the beam axis. This scaling is
to be compared with the $(a/\lambda)^4$ scaling for plane wave photon scattering (where
OAM exchange is absent). Since $w(0) \gtrsim \lambda$ by the uncertainty principle, we see that the absence of $a/\lambda$ suppression in OAM transfer is more then compensated for by $a/w(0)$ suppression, at least for an atom centered on the beam axis. 

Since $a /w(0) \ll 1$, using $w(0) = \sqrt{\lambda z_R/\pi}$ we necessarily have $a/z_R \ll 1$.
Under these conditions,  $ w(z) = w(0) ( 1 + z^2/z_R^2)^{1/2} \simeq w(0)$, 
so that from Eq.~(\ref{Mt2a}) one has,  
\begin{equation}
M \simeq  \frac{2p!}{\pi (p + |\ell  |)!}       \left(  \frac{(p + |\ell|)!}{p!  |\ell|!}  \right)^2    
 \int_0^\infty r^2 dr \int_0^\pi \sin(\theta) d\theta  \int_0^{2\pi} d\phi \   \  \phi_{N,L_f,M_f}^* (r,\theta,\phi)       
\nonumber\end{equation}
\begin{equation}\times
 \frac{1}{(w(0))^2} 
\left[\frac{ r}{w(0)}\right]^{2|\ell | }  |\sin \theta|^{2 |\ell|} 2^{|\ell|}  
e^{i 2 \ell \phi}    \phi_{N,L_i,M_i} (r,\theta,\phi)
\nonumber
\end{equation}
\begin{equation}
=   \frac{2p!}{\pi (p + |\ell  |)!} \ \  2^{|\ell|} \     \left(  \frac{(p + |\ell|)!}{p!  |\ell|!}  \right)^2 \frac{
\langle r^{2 |\ell|}\rangle_{N, L_i; N, L_f }}{(w(0))^{2 (|\ell| + 1)}}
   \langle \sin^{2 \ell}\theta\rangle_{N, L_i; N, L_f } \,,
\label{Mt2a1}
\end{equation} 
where $ \langle v \rangle_{N, L_i; N, L_f} = \int  \phi_{N,L_f,M_f = 0}^* (r,\theta,\phi)\, v(\vec{r}) \, 
\phi_{N,L_i,M_i = 0} (r,\theta,\phi) \ d \vec{r}$ is a matrix element of $v$,
and $M_f - M_i = 2 \ell$.  
We note, that to lowest order in this expansion, the matrix elements
of $r^{2 (|\ell| + 1)}$  and $\sin^{2 |\ell|}\theta$ are independent of each other since
the hydrogenic wave function is separable in $r$ and $\theta$.
If $L_i \neq L_f$, there can be a geometrical mismatch that may reduce the 
magnitude of the matrix element.
When $L_i = L_f = |\ell| = L$, we
obtain the simple and instructive expression, 
\begin{equation}
M \simeq  \frac{2p!}{\pi (p + |\ell  |)!} \ \  2^{|\ell|} \     \left(  \frac{(p + |\ell|)!}{p!  |\ell|!}  \right)^2 
\frac{1}{w^2(0)} \frac{\langle \rho^{2|\ell|}\rangle_{N, |\ell|; N, |\ell|}}{(w(0))^{2|\ell|}} \,,
\label{Mt2a1ell}
\end{equation}
where $\rho= r \sin \theta$ and $\langle \rho^{2|\ell|}\rangle \sim a^{2|\ell|}$, and $a$ is the target size.
Thus, the transition rate is relatively small when 
$a/w(0) \ll 1$ (and thus $a/z_R \ll 1$). 
It is also relevant to the transition rate that $a$
scales as the square of the principal quantum number, $N = |\ell| + 1$, 
so that $\langle r^{2 |\ell| }\rangle \sim N^{4(N-1)}  a_0^{2(N-1)}$, where $a_0$ is the Bohr radius. 

\section{Interpretation and discussion}

\subsection{Atomic targets}

A plane-wave photon may be characterized, apart from its direction of motion, by three parameters, ($\lambda, s, m_s$), where $s=1$ and $m_s = \pm 1$.  A free electron may also be characterized by three parameters, ($\Lambda, S, M_S$), where $S=\frac{1}{2}$ and $M_S = \pm \frac{1}{2}$ (here $\Lambda$ denotes the wavelength of the free electron).  A twisted (Laguerre-Gaussian profile) photon may be characterized, in addition to its direction of motion and waist width, by six parameters, ($\lambda, p, |\ell|, m_\ell, s, m_s$), where $m_\ell = \pm\ell$.  An atomic electron may be characterized by five parameters, ($N, L, M_L, S, M_S$), where $M_L$ may vary between $-L$ and $+L$ in integer steps.

The mathematical symmetry of a twisted photon differs from that of a plane-wave photon or Gaussian beam
due to the presence of the $e^{i \ell \phi}$ factor in the spatial distribution of the twisted photon.
In general, either through absorption/emission or scattering, a plane wave photon or Gaussian beam may transfer spin angular momentum to a hydrogen atom, whereas twisted photons can transfer orbital or spin angular momentum or both to a hydrogen atom. However, the transfer of $\ell$ and $s$ have a different mathematical symmetry, corresponding to different parity of the photon-atom system.  
This corresponds to the different symmetries of the electric vector field, $\vec{E}$, and magnetic pseudo-vector field, $\vec{B}$.
Changes in $\ell$ are caused by $\vec{E}$ at leading order, while $\vec{B}$ causes changes in $s$.   
For plane-wave photons, the matrix element for forward scattering is zero 
(see Eq.~(\ref{I})) due to the orthogonality of the atomic states for different $M$.  
It is to be noted that the expansion of Eq.~(\ref{znif}) for plane-wave photons in powers of $\vec{q} \cdot \vec{r}$ vanishes at zeroth order, while the 
term of the same order in Eq.~(\ref{Mt2p}) for twisted photons is non-zero. 
Thus the $e^{i \ell \phi}$ factor in OAM transfer effectively removes parity blocking 
present in SAM transfer
($\langle F|I\rangle$=0), analogous to restoration of right-handedness
in a reflection from a second mirror.

The symmetry of the Gouy phase around $z = 0$ (which we also take to be the center of the atom)
is essential to our result.
Specifically, if the Gouy phase depended on $\ell$ instead of $|\ell|$,
the  $\ell \to -\ell$ scattering matrix element in Eq.~(\ref{Mt2p}) would 
contain a phase factor, $e^{i 2 \ell \tan^{-1}({z}/{z_R})}$. 
This phase is odd in $z = r \cos\theta$ and would cause
the integration over $\theta$ in Eq.~(\ref{Mt2p}) to vanish, implying a forbidden OAM $\ell \rightarrow -\ell$ transition.

We also note that while $M_f - M_i = \Delta M = 2\ell$ is  required in Eq.~(\ref{Mt2p}),  
the special case $L_i=L_f=\ell$ is taken in Eq.~(\ref{Mt2a1ell})
for mathematical convenience only. In general, the restriction
on the electronic angular momentum in OAM transfer is only $L_i+L_f \ge 2|\ell|$.

For elastic scattering of twisted photons, there is transfer of $2\ell$ units of OAM with no change
in the energy of the photon.  Since $\hat{q}$ is perpendicular to the photon beam direction $\hat{k}$ at $\Theta = 0$,
the $2\ell \hbar$ angular momentum transfer is perpendicular to the direction of the photon beam. 
This transfer of $2\ell \hbar$ units of angular momentum occurs 
without an exchange of either energy or linear momentum; thus the process is force-free but not torque-free.

We have assumed in Sec.~\ref{sec:assumptions} that the hydrogen atom is at the center of the twisted photon vortex.
This was done because it enhances the symmetry of the photon-atom system.  
We have used this symmetry in both our derivations and our analysis.
If the atom is not at the center of the vortex, the transition rate may change.
This seems evident from the $\left[\sqrt{2} r\sin \theta /w(r \cos\theta)\right]^{2|\ell|}$
factor in Eq.~(\ref{Mt2p}), corresponding to a suppression in the density distribution of the twisted photon on the beam axis.
Since the atom is small compared to the beam width $w$, there is little overlap between the atom and the
beam when the atom is on the beam axis for $\ell \ne 0$.
This is reflected in Eq.~(\ref{Mt2a1ell}), where the matrix element goes rapidly to zero as $a/w(0) \rightarrow 0$.
It may be that moving the atom off the center of the vortex will change the strength of the interaction
with the photon beam.  

\subsection{Macroscopic targets}

If, for example, a properly oriented cubic crystal 
is used as a target, the rate of transfer of $2\ell$ units of OAM may be large if the outgoing beam is coherent.
It may be helpful to match the wavelength of the twisted photon to an integer times the lattice constant of the crystal
so that the contributions from each unit cell add coherently.
Various crystalline geometries may occur that lead to such coherence.

\subsection{Future directions and applications}

We note that the geometric size of the cross section for scattering of plane-wave 
visible photons from atoms is 
the Thompson cross section $8 \pi r_0^2/3 =6.3 \times 10^{-25}\, {\rm cm}^2 $,
where $r_0= \alpha^2 a_0$ is the classical electron radius.  Thus, cross sections for scattering of light from atoms are typically $\alpha^4 \sim 10^{-8}$ times smaller than those for scattering of charged particles (or atoms) from atoms, where the corresponding geometric cross section is $\sim \pi a_0^2$.  
The suppression of the cross section for plane-wave photons occurs because the expansion 
in $\vec{q} \cdot \vec{r}$ (of order $a/\lambda$) vanishes at zeroth order.  
For OAM transfer in twisted photon scattering, the series includes a zeroth-order term and the cross section is not suppressed.

By comparing the matrix element for OAM transfer in twisted beams with the matrix element for the scattering of a Gaussian beam on an atomic target, a properly normalized cross section for OAM transfer may be obtained~\cite{McGuire1,MC1}.
The Klein-Nishina effect, not included in this paper but often included in standard formulae
for Compton scattering, is small when $\lambda > 2.5 \times 10^{-12}$ m.
In our view, it is also interesting to reformulate this problem
using the particle picture, which yields probability amplitudes automatically normalized to unity. For scattering of protons (or ions or electrons) from atoms~\cite{McGuire2, MS},  
the scattering amplitude, $f(\vec{q})$ in the wave picture, 
is related to the transition probability amplitude, $a(\vec{b})$ in the particle picture, by 
$f(\vec{q}) =(M_r v/2 \pi)^2 \int  e^{ i \vec{q} \cdot \vec{b}} a(\vec{b}) d^2\vec{b}$,
where $M_r$ is the reduced mass 
of the projectile-atom system, $v$ is the projectile speed, and $\vec{b}$ is the impact
parameter.
For incident plane-wave protons, corresponding to straight-line classical trajectories, 
one then has $\vec{R}(t) = \vec{b} + \hat{z} vt$. 
We plan to investigate the application of this approach to the scattering of twisted light,
where in the classical limit $\vec{R}(t)$ corresponds to the  classical trajectory of the photon. Furthermore, we intend to explore the effect of relaxing the assumption made earlier, where the atom is restricted to be at the center of the twisted vortex.

Using the particle representation, it may be sensible, under appropriate physical conditions, 
to treat elastic forward scattering from atomic hydrogen as a coupling between two
degenerate qubits, or
two-state systems. 
A similar degenerate qubit has been analyzed by Shakov et al.~\cite{Shakovetal} for $2s-2p$ transitions   
in atomic hydrogen using plane-wave photons with wavelengths in the microwave regime. 
Their simple system yields simple algebraic solutions that can be used to determine
conditions under which the transition amplitude from one state to the other is either 1 or 0,
corresponding to the action integral being an integer multiple of $\pi \hbar/2$, or a phase rotation being a multiple of $\pi/2$ in the simplest case.  
Our case appears to be mathematically similar, with the action integral being, for example, proportional to
$\frac{\hbar^2}{2m}  \int \langle\vec{A}_f \cdot \vec{A}_i\rangle dt 
=\frac{\hbar T}{2m}  \langle (\vec{A}_f \cdot \vec{A}_i) \rangle$, 
where $T$ is the duration of a twisted photon pulse that is suddenly turned on and then off.  
The transition matrix element of interest for a twisted photon elastically scattering into the 
forward direction  is proportional to that given  in Eq.~(\ref{Mt2p}). 
This twisted photon-atom system corresponds to a two-qubit system that can be controlled.  
Experimental evidence for quantum control has been available for several years~\cite{Bucks}. 
In this way the transfer of $2 \ell \hbar$ of angular momentum using twisted photons could have applications in quantum information.  

We speculate that, in principle, it might be possible to modulate the twisted beam
to transfer information to different atoms (or locations in a macroscopic target).
However, we point out that such a modification would most likely be useful
when the beam waist $w(0)$ is not very large compared to the target size, $a$. 
Such an application using an atomic target seems remote at this time. 

\section{Summary}

In this paper we have derived explicit expressions for the matrix element for a twisted 
(Laguerre-Gaussian) photon interacting with a hydrogen atom.
While a pure Gaussian or plane-wave photon caries an intrinsic spin, 
$\vec{s}$, a twisted photon carries an an additional orbital angular momentum, 
$\vec{\ell}$, which can also be transferred to a hydrogen atom.

Symmetry plays an important role in this matrix element.  
The factor of $e^{i \ell \phi}$ present for a twisted photon, 
but not in a plane-wave photon, provides a different mathematical symmetry
in the matrix element for the transfer of OAM to an atomic electron
compared to transfer of SAM.
Specifically, the matrix element for OAM transfer has a series expansion that includes
a term of zeroth order in
$\vec{q} \cdot \vec{r}$, whereas in the corresponding series for plane wave photon scattering,
such a term vanishes.

Geometric symmetry also plays a role.  Unless an external axis is imposed on the atom (e.g. by an electric or magnetic field), the atom is initially symmetric.  
Twisted photons have a cylindrical symmetry.
Three different geometric length scales appear in our formulation:  
$a$, the size of the target; $\lambda$, the photon wavelength; 
and $z_R$, the Rayleigh range that determines the vortex geometry of the twisted photon. 
Each may be adjusted physically and may thus be used to vary reaction rates with twisted photons.  In the case of a hydrogen atom, the target size, $a$, may be increased 
from the Bohr radius by preparing the target in a Rydberg state. 
The expressions of Allen~\cite{AllenBSW}, which we use, are obtained in the paraxial approximation, and require that $\lambda \ll z_R$.

Forward scattering introduces a symmetry not present when the photon is scattered 
into a finite angle.  
Elastic scattering provides an additional symmetry in which $\hat{q}$ is perpendicular to $\hat{k}$ in the forward direction, where transfer of OAM appears to be relatively large.
For elastic scattering of twisted photons in the forward direction,
we have presented simple expressions for the matrix elements with explicit scaling in the 
hydrogenic quantum numbers $N$ and $L$ for some special limiting cases. When the atomic size $a$ is small compared with the beam waist $w(0) \sim \sqrt{\lambda z_R}$, the transition matrix element for OAM transfer can become quite small.
In addition the matrix element varies with the radial index $p$ of the twisted photon (defining
the size of the radial distribution) and the principal quantum number $N$ of the atom 
(defining the size of the atom) as well as with the photon winding number $\ell$.
Thus the transition rates may be controlled by physically adjusting one or more of the spatial scales
$z_R$, $\lambda$, and $a$, and the quantum numbers $p$, $\ell$, $N$, and $L$.

\acknowledgments
We thank D. B. Uskov for discussion related to applications of twisted photons.
JHM acknowledges useful discussions with Alex Clark and V. J. McGuire
related to mathematical symmetry.
This work was supported in part by the NSF under Grant No. PHY-1005709.

\end{document}